\documentclass[a4paper, USenglish, cleveref, autoref]{lipics-v2021}

\hideLIPIcs  

\usepackage{xspace}
\usepackage{svg}
\usepackage{adjustbox}

\NewDocumentCommand\TableauxRocq{}{\textsf{TableauxRocq}\xspace}
\NewDocumentCommand\isabelle{}{\textsf{Isabelle}\xspace}
\NewDocumentCommand\Rocq{}{\textsf{Rocq}\xspace}
\NewDocumentCommand\goeland{}{\textsf{Goeland}\xspace}
\NewDocumentCommand\lp{}{\textsf{LambdaPi}\xspace}
\NewDocumentCommand\lisa{}{\textsf{Lisa}\xspace}
\NewDocumentCommand\tptp{}{\textsf{TPTP}\xspace}
\NewDocumentCommand\sctptp{}{\textsf{SC-TPTP}\xspace}
\NewDocumentCommand\git{}{\textsf{GitHub}\xspace}
\Crefname{section}{Sec.\@}{Sec.\@}
\Crefname{figure}{Fig.\@}{Fig.\@}
\NewDocumentCommand\shepherd{m}
  {#1}

\title{\texorpdfstring{\TableauxRocq: A Deep Embedding of Free-Variable Tableaux in \Rocq}{TableauxRocq: A Deep Embedding of Free-Variable Tableaux in Rocq}}

\author{Johann Rosain}{\textsf{ENS de Lyon}, Lyon, France}{johann.rosain@ens-lyon.org}{https://orcid.org/0000-0003-1719-2654}{}
\author{Julie Cailler}{\textsf{University of Lorraine}, \textsf{CNRS}, \textsf{Inria}, \textsf{LORIA}, Nancy, France} {julie.cailler@inria.fr}{https://orcid.org/0000-0002-6665-8089}{}

\authorrunning{J.Rosain and J. Cailler} 

\Copyright{Johann Rosain and Julie Cailler} 



\category{} 

\relatedversion{} 

\supplement{\shepherd{The \TableauxRocq version presented in this paper together with the modified
  version of \goeland and the scripts to run the experiments are available on
  Zenodo~\cite{zenodo_sm}. The \TableauxRocq library is available online at \url{https://github.com/jrosain/TableauxRocq}.}}

\acknowledgements{We would like to thank the anonymous reviewers for their
    valuable and insightful comments on the paper.}

\nolinenumbers 

\EventEditors{Ekaterina Komendantskaya and Tobias Nipkow}
\EventNoEds{2}
\EventLongTitle{17th International Conference on Interactive Theorem Proving (ITP 2026)}
\EventShortTitle{ITP 2026}
\EventAcronym{ITP}
\EventYear{2026}
\EventDate{July 26--29, 2026}
\EventLocation{Lisbon, Portugal}
\EventLogo{}
\SeriesVolume{382}
\ArticleNo{13}

\usepackage{acro}
\usepackage{subcaption}
\usepackage{mathpartir}

\definecolor{darkblue}{rgb}{0.0, 0.0, 0.55}
\definecolor{darkgreen}{rgb}{0.0, 0.55, 0.13}
\definecolor{darkred}{rgb}{0.55, 0.0, 0.0}
\definecolor{darkviolet}{rgb}{0.58, 0.0, 0.83}
\definecolor{lightblue}{rgb}{0.68, 0.85, 0.9}
\usepackage{lstrocq}

\usepackage{xcolor}
\usepackage{soul}
\definecolor{hlcolor}{HTML}{
    E3D7D4
}
\sethlcolor{hlcolor}

\usepackage{tikz}
\usepackage{pgfplots}
\pgfplotsset{compat=1.18}

\NewDocumentCommand{\eg}{}{\emph{e.g.,}\xspace}
\NewDocumentCommand{\ie}{}{\emph{i.e.,}\xspace}

\NewDocumentCommand{\defeq}{}{\ensuremath{\;\triangleq\;}}
\NewDocumentCommand{\stdlib}{}{\texttt{Stdlib}\xspace}

\NewDocumentCommand{\interp}{m}{\ensuremath{[\![ #1 ]\!]}}
\NewDocumentCommand{\para}{m}{\subparagraph*{#1.}}

\DeclareAcronym{atps}{
  short=ATPs,
  long=Automated Theorem Provers,
}
\DeclareAcronym{stdlib}{
  short=\stdlib,
  long=\Rocq's standard library,
}
\DeclareAcronym{fol}{
  short=FOL,
  long=first-order logic,
}

\NewDocumentCommand{\mynote}{%
  m 
  m 
  +O{red} 
}{\fbox{\bfseries\sffamily\scriptsize\textcolor{#3}{#1}}
    {\textcolor{#3}{\small$\blacktriangleright$\textsf{\emph{#2}}$\blacktriangleleft$}}\xspace}

\NewDocumentCommand{\rocqm}{m}{\texttt{\small#1}}
\NewDocumentCommand{\interpret}{
  m 
  m 
  m 
  m 
}{\interp{\rocqm{#1}\ \rocqm{\#}\ #2\ \rocqm{\#}\ #3 \models #4}}
\NewDocumentCommand{\variableOpening}{
  m 
  m 
  m 
}{\ensuremath{\rocqm{#1\{#2} \to \rocqm{#3\}}}}
\NewDocumentCommand{\validunder}{
  m 
  m 
}{\ensuremath{#1 \models #2}}

\newcommand{\setRocqFilename}[1]{\def\filename{#1}}
\newcommand{\setBaseUrl}[1]{\def\baseurl#1}
\setRocqFilename{}
\setBaseUrl{{https://jrosain.github.io/TableauxRocq/v0.1-ITP26/}}

\usepackage{nameref}

\makeatletter
\DeclareDocumentCommand \galinferrule { o m m }{%
  \IfValueTF{#1}%
  {\mpr@inferstar*[right=#1]{#2}{#3}%
   \def\@currentlabelname{\textsc{#1}}}%
  {\mpr@inferstar*{#2}{#3}}}

\newcommand{\insertlabel}[1]{\label{rocq:#1}}
\usepackage{mfirstuc}
\usepackage{underscore}
\newcommand{\rocqlink}[1]{\href{\baseurl\filename.html\##1}{\protect\raisebox{-.7ex}{\protect\includegraphics[height=11pt]{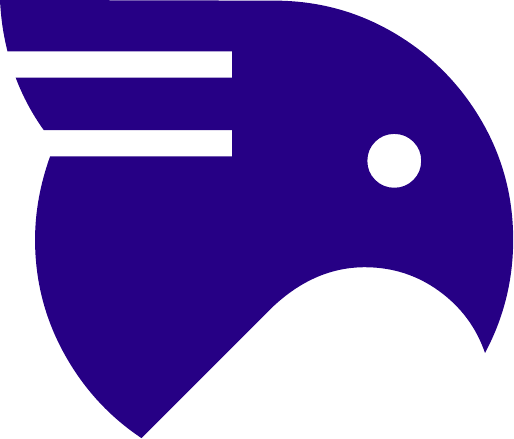}}}}

\newtheoremstyle{rocqtheorem}%
  {}{}%
  {}{}
  {\textcolor{darkgray}{$\blacktriangleright$}\nobreakspace\sffamily\bfseries}{.}
  {.5em}{\thmname{#1}\thmnumber{ #2}\normalfont\sffamily\thmnote{ (#3\xspace\rocqlink{\theoremname})}}

\theoremstyle{rocqtheorem}

\newcommand{\genFancyTheoremEnvironment}[1]{%
    \expandafter\providecommand\csname #1autorefname\endcsname{\capitalisewords{#1}}
    \expandafter\newcommand\csname #1Auxautorefname\endcsname{\csname #1autorefname\endcsname}
    \expandafter\newcommand\csname #1AuxRocqautorefname\endcsname{\csname #1autorefname\endcsname}

    \@ifpackageloaded{cleveref}{
      \crefformat{#1Aux}{{\capitalisewords{#1}}~##2##1##3}
      \Crefformat{#1Aux}{{\capitalisewords{#1}}~##2##1##3}
    }{}

    \newtheorem{#1Aux}[dummy]{\capitalisewords{#1}}

    \NewDocumentEnvironment{rocq#1}{oo}%
    {%
      \IfValueT{##2}{\def\theoremname{##2}}%
      \IfValueTF{##1}{\ifx&##1&%
        \begin{#1Aux}[]%
          \else\begin{#1Aux}[##1]\fi}%
          {\begin{#1Aux}[]}
            \IfValueT{##2}{\insertlabel{##2}}%
          }%
          {\end{#1Aux}\def\theoremname{}}}

\genFancyTheoremEnvironment{lemma}
\genFancyTheoremEnvironment{requirement}
\genFancyTheoremEnvironment{definition}
\genFancyTheoremEnvironment{theorem}
\genFancyTheoremEnvironment{corollary}
\makeatother

\theoremstyle{definition}
\newtheorem*{requirement*}{Requirement}

\NewDocumentCommand{\expands}{
  m 
  m 
}{\ensuremath{#1 \rhd #2}}

\NewDocumentCommand{\de}{}{deep embedding\xspace}

\NewDocumentCommand{\dummyName}{}{\phantom{a}\hspace{-8pt}}

\begin{document}

\maketitle

\begin{abstract}
  The free-variable tableau method has been widely used in order to automate proofs in
  multiple kinds of logics. Many automated theorem provers rely on this approach, either
  because it is the only available method---\eg{} in certain modal logics---or because it
  facilitates the generation of proof certificates. However, as far as the authors know,
  its results have never been formalized in a proof assistant.  In this paper, we present
  \TableauxRocq, a \de of free-variable first-order tableaux in the \Rocq prover. The
  formalized calculus is proved sound and provides a modular Skolemization system that
  enables the use of Skolemization-based optimizations. Moreover, we show how
  \TableauxRocq can be used as a certifier for automated theorem provers by adapting the
  \goeland prover---that can already output \Rocq terms---to output
  proofs in the \TableauxRocq format. By using the power of reflection, thereby providing
  a fully certified proof checker for free, we show that \goeland's exported
    \Rocq terms and \TableauxRocq's proof certificates can be checked in a similar time
    frame without proof optimizations, and that the latter has strictly better
    performances in presence of Skolemization-related optimizations.
\end{abstract}

\section{Introduction}
\label{sec:introduction}

The free-variable tableau method~\cite{MFit96} is a proof-search procedure originally
designed for \ac{fol}. This technique, in contrast to clausification-based
approaches~\cite{andrews1981theorem,bibel82atp,DBLP:journals/logcom/BachmairG94,robinson1965machine,nieuwenhuis2001paramodulation},
allows the search for proofs to operate directly from the base formula, thus providing a
more trusted proof output. Indeed, by avoiding complex transformations, 
these techniques provide more human-readable derivations, producing proofs that are in turn easier to (machine-)check.
Free-variable tableaux have then been extended to first-order logic with theory reasoning~\cite{beckert1996incremental,delahaye2013zenon} 
such as arithmetic~\cite{rummer2008constraint,bury2015integrating} or rank-$1$ polymorphism~\cite{burel2020first}, and to other logics, \eg
modal logics~\cite{DBLP:journals/sLogica/BeckertG01}, intuitionistic
logics~\cite{DBLP:conf/iis/KonevL06}, and higher-order
logic~\cite{DBLP:conf/tableaux/Kohlhase95}. Due to their design as proof-search
procedures, the various tableau calculi include many optimizations, making their soundness and completeness proofs rely on non-trivial arguments. In practice, these optimizations
lead to complex implementations in
\ac{atps}~\cite{delahaye2007zenon,cailler2022goeland,rummer2008constraint,beckert1995leantap,beckert1996tableau,oppacher1988harp},
which in turn makes it difficult to fully trust their answers. Consequently,
recent efforts in the \ac{atps} community have focused on certifying
the output of their
tools~\cite{delahaye2007zenon,cauderlier2015checking,LPAR2024:Generic_Deskolemization_Strategy,komel2025case,lachnitt2025improving,rawson2025ground}, 
relying mostly on \emph{syntactic} methods\shepherd{, \ie by translating their proofs into a given format, with potential transformations and elaboration phases.}

In this paper, we advocate the use of \emph{semantic} methods based on a \de
in proof assistants. Conversely to syntactic methods, semantic ones propose
to transfer the development effort from the \ac{atps}' translation mechanism to the certification part by formalizing the underlying theory, 
thus allowing to readily use proofs generated by \ac{atps} as-is.
Semantic methods offer several advantages over syntactic ones. For
instance, they make it possible to formally establish the soundness and completeness of
\emph{optimized} versions of the calculi, which in turn enable \ac{atps} to output proofs
with no further syntactic manipulations. This can result in a significant speedup in both
proof generation and proof checking time, as some optimizations---such as Skolemization in
first-order logic---are known to cause an explosion in the size of the proof
certificates~\cite{baaz2012deskolemization}. Moreover, semantic methods shift confidence in 
the certificate from a \emph{trusted code base} to a \emph{trusted theory base} paradigm.
\shepherd{
In syntactic 
approaches, beyond trusting the proof assistant's kernel, one must also trust the 
correctness of the proof-translation pipeline---a potentially complex mechanism 
that manipulates proof terms and may introduce errors. In semantic approaches, this 
additional trust is instead placed in the formalized theory itself: a set of 
mathematical definitions and lemmas that can be independently inspected and 
validated by a domain expert, without requiring any understanding of the 
implementation.
}

\shepherd{
Semantic approaches also allow for the formal development of
soundness and completeness proofs for actual proof-search procedures, which is a topic
that, while being well studied for propositional logic~\cite{fleury2018verified,blanchette2018verified,fleury2020formalization,holldobler2014generic}, is more scattered for \ac{fol}~\cite{DBLP:journals/jar/Schlichtkrull18} and as far as the authors know, is missing for tableaux.}

\para{Contributions} Our contributions are threefold. First, we develop \TableauxRocq, a
\de of the first-order free-variable tableau calculus in \Rocq, presented in
\cref{sec:formalized-calculus}, that is modular w.r.t.\ the Skolemization method used. We
provide a generic specification of a Skolemization step that allows us to prove the
soundness of the tableau calculus w.r.t.\ the usual classical first-order semantics.
Second, in \cref{sec:proof-checker}, we implement an algorithm that decides whether a proof
is a valid tableau proof and prove it sound w.r.t.\ the specification. Finally, we adapt
\goeland~\cite{cailler2022goeland}, a first-order tableau-based theorem prover supporting
multiple Skolemization strategies to output proofs in the \TableauxRocq format. 
We also report on initial benchmarks validating our approach in \cref{sec:comparison-benchs}.

\setRocqFilename{index}
Throughout the paper, we provide links to an HTML-rendered version of the \Rocq
formalization, signaled by a small \Rocq logo \rocqlink{}. The formalization is available
on \git.\footnote{At \url{https://github.com/jrosain/TableauxRocq}.} 
For ease of writing, everything in the code is in ASCII, but both in the HTML version and
the paper, we render some notations in UTF-8 characters.


\section{Formalized Free-Variable First-Order Tableau Calculus}
\label{sec:formalized-calculus}

In this section, we describe the core of the \TableauxRocq library. It contains an
implementation of the free-variable first-order tableau calculus using a minimal syntax,
which makes it possible to avoid redundant cases in proofs while not limiting
expressiveness. As is usually done in syntaxes with binders, we use a specific
representation of variables, presented in \cref{sec:locally-nameless-repr}. Based on this
representation, we define a minimal syntax and semantics for \ac{fol} in
\cref{sec:syntax-semantics}.  We then introduce an axiomatization of Skolemization in
\cref{sec:skolemization}, which enables to formulate a modular tableau calculus in
\cref{sec:tableaux-proofs} and prove its soundness in \cref{sec:soundness}.

\subsection{A Locally Nameless Representation of Variables}
\label{sec:locally-nameless-repr}

Systems with binders (\eg{} $\forall$ and $\exists$ in \ac{fol}) are tricky to formalize.  In
pen-and-paper proofs, definitions such as $\alpha$-conversion and substitution are often
treated informally, relying on the reader's intuition to fill in the gaps.  However,
directly formalizing these definitions in a proof assistant incurs significant
overhead. Hence, a common solution is to use De Bruijn
indices~\cite{DBLP:conf/lambda/deBruijn72}, a representation in which binders do not
explicitly name their variables. Instead, variables are replaced by natural numbers
indicating the number of binders between the variable and its corresponding binder. This
approach automatically handles $\alpha$-conversion and allows for a straightforward definition
of substitution. Consequently, in dependent-type based provers, most systems
with binders are formalized using De Bruijn
indices~\cite{DBLP:conf/cpp/AdjedjLMPP24,DBLP:journals/jacm/SozeauFLNTW25,forster2020coq,DBLP:journals/pacmpl/FioreS22,DBLP:journals/pacmpl/Winterhalter24,DBLP:journals/jar/TiroreBC25}.

In \TableauxRocq, we adopt a locally nameless
representation~\cite{DBLP:journals/jar/Chargueraud12} for managing variables.  This
approach keeps the advantages of the De Bruijn representation---by implementing bound
variables with natural numbers---while providing more flexibility in handling free
variables. Because tableau proofs involve a list of formulas rather than a single one,
maintaining the De Bruijn indices of free variables synchronized would require substantial
effort, which the locally nameless representation allows us to avoid.

\setRocqFilename{Tableaux.Prelude.Atoms}
\begin{lstlisting}[caption={\Rocq Implementation of the Typeclass of Atoms \rocqlink{Atom}}, label={listing:class-atoms}, language=rocq, float=tbp]
Record Atom :=
  { atom :> Type
  ; eqb_atom :: EqBool atom
  ; set_atom : set atom }
\end{lstlisting}

In our use case, the type of free variables cannot be treated as a bare type; it needs
additional structure. In \TableauxRocq, this structure is given by a type that
  we call \rocqe|Atom|: a type that has (i) a way to decide equality between its elements
  and (ii) the ability to collect them in a set-like structure, for example when
computing the set of variables appearing in a formula. The implementation of
  \rocqe|Atom| is given in \cref{listing:class-atoms}. Note that, since our aim is to
write executable algorithms, an \rocqe|Atom| satisfies the \rocqe|EqBool| class---a boolean
equality \rocqe|eqb| that reflects the propositional one---which is equivalent to point
(i). The \rocqe|atom| field, \ie the \emph{atomic} type, allows us to provide
\emph{structures} for \eg natural numbers or strings, and to write %
\rocqe|Check nat : Atom|. This makes it possible to pass \rocqe|nat| directly to a
function expecting an \rocqe|Atom|, without explicitly providing its \rocqe|Atom|
instance. The token \rocqe|:>| indicates an implicit coercion from an \rocqe|Atom| to its
carrier type, allowing one to write \rocqe|x : A| instead of
\rocqe|x : atom A| for \rocqe|A : Atom|. Similarly, the token \rocqe|::| automatically
declares an instance of the typeclass \rocqe|EqBool|, enabling one to write
\rocqe|eqb x y| instead of \rocqe|eqb (eqb_atom A) x y|.

We also provide type classes and definitions for elementary locally nameless operations,
allowing for the same notations to be shared across different structures implementing
locally nameless variables. In the following, \rocqe|t| and \rocqe|u| denote terms and
\rocqe|F| a formula.

\setRocqFilename{Tableaux.Prelude.LocallyNamelessClasses}
\begin{itemize}
\item variable opening (\rocqlink{Opening}), denoted \variableOpening{t}{n}{u}, which replaces the $n^{\text{th}}$ bound variable in \rocqe|t| with \rocqe|u|,
\item bound variables computation (\rocqlink{BV}), denoted \rocqe|bv F| or \rocqe|bv t| giving the bound variables of \rocqe|F| or \rocqe|t|,
\item free variables computation (\rocqlink{FV}), denoted \rocqe|fv|, the counterpart of \rocqe|bv| for free variables,
\item local closure (\rocqlink{isLocallyClosed}), denoted \rocqe|isLocallyClosed|, which is true if there are no \emph{bound} variables in a term or in a formula,
\item closure (\rocqlink{isClosed}), denoted \rocqe|isClosed|, which is true if there are no \emph{free} variables in a term or in a formula, and
\item substitution (\rocqlink{Substitution}), denoted \rocqe|x@[sigma]|, which maps atoms to locally closed terms.
\end{itemize}

\subsection{A Minimal Syntax and Semantics}
\label{sec:syntax-semantics}
\setRocqFilename{Tableaux.Syntax}

\begin{figure}
  \centering
  \fbox{
    \parbox{0.97\linewidth}{
      \begin{align*}
        \text{(terms \rocqlink{Term})} & &t \quad &::= \quad n \mid x \mid f(t_1, \ldots, t_n) & n \in \mathbb{N}, x \in \mathcal{V} , f \in \mathcal{F}\\
        \text{(formulas \rocqlink{Form})} & &F, G \quad &::= \quad \bot \mid P(t_1, \ldots, t_n) \mid \neg F \mid F \vee G \mid \forall F & P \in \mathcal{P}
      \end{align*}
    }
  }
  \caption{Minimal Locally Nameless First-Order Syntax}
  \label{fig:minimal-syntax}
\end{figure}

\TableauxRocq implements the \ac{fol} locally nameless syntax of
\cref{fig:minimal-syntax}.  This syntax is minimal: it includes terms (integers,
variables, and functions) and formulas (bottom~$\bot$, predicates, negation~$\neg$,
disjunction~$\lor$, and universal quantification~$\forall$).  In this formalism, the formula
$\forall x\,(\forall y\,R(a, Z, x, y))$ is represented in locally nameless style as
$\forall(\forall(R(a, Z, 1, 0)))$, where $a$ is a constant, $Z$ is a free variable, and the bound
variables are natural numbers indicating how many $\forall$s must be discarded to reach the
binder they refer to.  Moreover, the sets \rocqe|var|, \rocqe|func|, and
\rocqe|pred|---which respectively represent free variables, function symbols, and predicate
symbols---are implemented as \rocqe|Atom|s that are (implicit) parameters of the syntax,
providing greater flexibility to the user. This design also allows us to show, for
example, that terms and formulas have a boolean equality---\ie{} we have the typeclass
instances \rocqe|EqBool Term| and \rocqe|EqBool Form|. Substitution and variable opening
can then be straightforwardly defined, making it possible to prove the expected
commutation lemmas between these operations.

\setRocqFilename{Tableaux.Semantics}
\begin{figure}
  \centering
  \fbox{
    \parbox{0.97\linewidth}{
      \begin{subfigure}{\linewidth}
        \begin{align*}
        \interpret{M}{\rho}{\sigma}{n} &\defeq \rho[n] \\
        \interpret{M}{\rho}{\sigma}{\rocqm{x}} &\defeq \sigma(\rocqm{x}) \\
        \interpret{M}{\rho}{\sigma}{\rocqm{f}(t_1, \ldots, t_n)} &\defeq \rocqm{M(f)}(\interpret{M}{\rho}{\sigma}{t_1}, \ldots, \interpret{M}{\rho}{\sigma}{t_n})
        \end{align*}
        \caption{Interpretation of Terms \rocqlink{interpret_term}}
      \end{subfigure}
      \begin{subfigure}{\linewidth}
        \begin{align*}
        \interpret{M}{\rho}{\sigma}{\rocqm{P}(t_1, \ldots, t_n)} &\defeq \rocqm{M(P)}(\interpret{M}{\rho}{\sigma}{t_1}, \ldots, \interpret{M}{\rho}{\sigma}{t_n}) \\
        \interpret{M}{\rho}{\sigma}{\rocqm{Neg}\ F} &\defeq \neg \interpret{M}{\rho}{\sigma}{F} \\
        \interpret{M}{\rho}{\sigma}{\rocqm{Or}\ F_1\ F_2} &\defeq \interpret{M}{\rho}{\sigma}{F_1} \lor \interpret{M}{\rho}{\sigma}{F_2} \\
        \interpret{M}{\rho}{\sigma}{\rocqm{All}\ F} &\defeq \forall (\rocqm{x} : \rocqm{M}),\ \interpret{M}{\rocqm{x ::}\ \rho}{\sigma}{F}
        \end{align*}
        \caption{Interpretation of Formulas \rocqlink{interpret_form}}
      \end{subfigure}
    }
  }
  \caption{Minimal Locally Nameless First-Order Semantics}
  \label{fig:minimal-semantics}
\end{figure}

\para{Semantics} The semantics implemented in \TableauxRocq follows the ideas of the
syntax: standard first-order semantics generalized over the types \rocqe|pred|,
\rocqe|func| and \rocqe|var|, as shown in \cref{fig:minimal-semantics}. Because the syntax
involves both bound and free variables, the interpretation of terms and formulas differs
slightly from the usual interpretation functions: we maintain both a bound- and a
free-variable environment. The former, which we mainly denote $\rho$, is represented as a
list of elements of the domain, while the latter, that is often denoted by $\sigma$ or
$\mu$, is implemented as a function. To allow for consistent notation across
interpretations, \TableauxRocq implements the interpretation function as a typeclass and
provides instances for terms and formulas.

We denote by \rocqe@[[ M # rho # sigma |- N ]]@ the interpretation of \rocqe|N| (which can
be \eg a term, a formula, a list of formulas, \ldots) in the model \rocqe|M|, with the
bound-variable environment \rocqe|rho| and the free-variable environment
\rocqe|sigma|. The usual lemmas about commutation between (i) variable opening and the
bound-variable environment (\rocqlink{term_env_inst_commutes},
\rocqlink{form_env_inst_commutes}), and (ii) substitution and the free-variable
environment (\rocqlink{subst_commutes_with_env_terms},
\rocqlink{subst_commutes_with_env_forms}) follow naturally.

A formula \rocqe|F| is \emph{satisfiable} (\rocqlink{is_satisfiable}) if there
exists a model \rocqe|M| such that for every free-variable environment $\sigma$,
\interpret{\rocqm{M}}{\cdot}{\sigma}{\rocqm{F}}, and \emph{valid} if for every model \rocqe|M|,
\interpret{\rocqm{M}}{\cdot}{\cdot}{\rocqm{F}}. In particular, we denote
$\validunder{\Gamma}{\rocqm{F}}$ if \rocqe|Or (Neg Gamma) F| is valid (where a context is
coerced into a formula by taking the conjunction of its elements).

\subsection{Axiomatizing Skolemization}
\label{sec:skolemization}

\setRocqFilename{Tableaux.Skolemization}
\begin{figure}[t]
  \fbox{
  \begin{mathpar}
    \inferrule{\neg (\forall F)}{\neg F \rocqm{\{0 }\to\rocqm{ sko(}X_1, \ldots, X_n\rocqm{)\}}}\delta
    \end{mathpar}}
  \caption{General Shape of a Skolemization Rule}
  \label{fig:shape-of-sko-rule}
\end{figure}

In \ac{fol} free-variable tableaux, the Skolemization rule applied to an
existentially quantified formula, whose general shape is presented in
\cref{fig:shape-of-sko-rule}, has many variations~\cite{MFit96,
  hahnle1994delta+,beckert1993delta++,baaz1995delta*,cantone98delta**,giese1999deltaepsilon}.
This rule instantiates the first bound variable of $F$ with the Skolem function returned by the \rocqe|sko| meta-function, which gives a Skolem symbol and selects a subset of the free variables occurring in the branch.

For instance, in the earliest version of the rule given by Fitting~\cite{MFit96}, called
\emph{outer Skolemization}, \rocqe|sko| returns a function symbol that is fresh (\ie not
already appearing in the tableau) and parameterized by all free variables of the
branch. Later, H{\"{a}}hnle and Schmitt~\cite{hahnle1994delta+} observed that
parameterizing only by the free variables occurring in the Skolemized formula suffices to
retain soundness. This method is known as \emph{inner Skolemization}. Moreover, they
showed that there exists a class of formulas $(\Phi_n)_{n \in \mathbb{N}}$ such that, if
$b(n)$ (resp. $b^+(n)$) denotes the number of branches in a proof of $\Phi_n$ using outer
(resp.\ inner) Skolemization, then $b^n = O(2^{b^+(n)})$.

To account for the different (un)known Skolemization strategies, we
formalize Skolemization as an operation specified by the following data
(\rocqlink{SkolemizationData}):
\begin{itemize}
\item a Skolemization record (\rocqe|sko_record|), explained in the last paragraph of the subsection,
\item a boolean-valued predicate
  \rocqe|is_sko : Term -> Form -> sko_record -> set_atom var -> set_atom func -> bool|,
  that checks whether a term is a valid Skolemized symbol of a formula given the free
  variables of the branch and the function symbols of the whole tableau,
\item a \rocqe|symbol| function that returns the function symbol of any valid Skolemized
  term $t$, and
\item an \rocqe|args| function that returns the arguments of any valid Skolemized term $t$.
\end{itemize}

This data must satisfy four requirements. Since the fourth one will only become relevant later in proofs, we first provide some intuition and defer the detailed explanation to \cref{sec:soundness}.

\begin{rocqrequirement}[\dummyName][is_func]
  For every term \rocqe|t|, if \rocqe|t| is a valid Skolemized term, then
  \rocqe|t = Fun (symbol t) (args t)|.
\end{rocqrequirement}

\begin{rocqrequirement}[\dummyName][args_sound]
  For every term \rocqe|t|, if \rocqe|t| is a valid Skolemized term, then the arguments of
  \rocqe|t| must all be free variables.
\end{rocqrequirement}

\begin{rocqrequirement}[\dummyName][is_sko_consistent]
  Every term \rocqe|t| that is a valid Skolemized term under a context with a set of
  function symbols \rocqe|s| is also a valid Skolemized term under a context with any
  subset of \rocqe|s| as function symbols.
\end{rocqrequirement}

\begin{rocqrequirement}[\dummyName][is_sko_sound]\label{requirement:sko-sound}
    For every term \rocqe|t| and model \rocqe|M|, if \rocqe|t| is a valid Skolemized term
    w.r.t.\ a formula \rocqe|Neg (All F)|, then there exists a model \rocqe|M|$'$ such that
    (i) for every formula \rocqe|G|, if \rocqe|G| is satisfied in \rocqe|M|, then it is
    also satisfied in \rocqe|M|$'$, and (ii) if \rocqe|Neg (All F)| is satisfied by
    \rocqe|M|, then \rocqe|Neg F{0 \to t}| is satisfied by \rocqe|M|$'$.
\end{rocqrequirement}
The first three requirements ensures that the predicate \rocqe|is_sko| is well-formed. The last one is the key ingredient for proving the soundness of the tableau method.

\para{Skolemization Record (\rocqlink{SkoRecord})} Skolemization records appear starting
from the \emph{pre-inner} Skolemization variant~\cite{beckert1993delta++}. This version of
the rule makes the generated Skolem symbol formula-dependent, \ie if $F$ is a formula to
which a $\delta$-rule is applied twice, the same Skolem symbol is produced both times. It is
therefore necessary to keep track of the formulas with which Skolem symbols are
associated. For example, for the outer and inner Skolemization rules, it suffices to
instantiate the Skolemization record as a set, since in these cases only freshness needs
to be ensured.

\setRocqFilename{Tableaux.Proofs}
\subsection{Tableaux Proofs}
\label{sec:tableaux-proofs}

\begin{figure}[t]
  \centering
  \fbox{
  \parbox{0.97\linewidth}{
    \begin{mathpar}
    \inferrule{\bot}{\odot}\odot_{\bot} \quad
    \inferrule{P, \neg Q}{\sigma \\ \sigma(P) = \sigma(Q)}\odot_{\sigma} \quad
    \inferrule{\neg (\neg F)}{F}\alpha_{\neg\neg} \quad
    \inferrule{\neg (F_1 \lor F_2)}{\neg F_1, \neg F_2}\alpha_{\neg\lor} \quad
    \inferrule{F_1 \lor F_2}{F_1 \\ F_2}\beta_{\lor} \\
    \inferrule{\forall F}{F\rocqm{\{0 }\to\rocqm{ }X\rocqm{\}}}\gamma_{\forall} \quad
    \inferrule{\neg (\forall F)}{\neg F \rocqm{\{0 }\to\rocqm{ sko(}X_1, \ldots, X_n\rocqm{)\}}}\delta_{\neg\forall}
  \end{mathpar}}}
  \caption[tableau-rules]{Minimal First-Order Free-Variable Tableau Rules}
  \label{fig:tableau-rules}
\end{figure}

The first-order tableau method is a forward-reasoning technique presented as an
  upside-down proof tree \shepherd{with root at top}, as illustrated in \cref{fig:tableau-rules}. It is composed of
  unary ($\alpha$, $\gamma$ and $\delta$) and binary ($\beta$) expansion rules, extending a branch with at
  most two formulas in the unary case and dividing a branch into two distinct ones in the
  binary case. $\odot$-rules are closure rules and can be applied on a branch whenever there is either a
  trivial contradiction, or if there are two formulas $P$ and $\neg Q$ and a substitution
  $\sigma$ making $\sigma(P)$ and $\sigma(Q)$ definitionally (\ie syntactically) equal. %
A tableau is said \emph{closed} if all its leaves are instances of closure rules. In that
case, since a tableau proceeds by \emph{refutation}, the negation of the root formula is
universally valid.

Our formalization mimics this forward-reasoning technique in three stages. First, we define a \rocqe|TableauTree| (\rocqlink{TableauTree}) as a binary tree whose nodes are labeled with lists of formulas.
Intuitively, each label corresponds to the formulas resulting from the application of a tableau rule. 
A \emph{branch} (\rocqlink{Branch}) is a list of \emph{branching steps} \rocqe|Left| or \rocqe|Right|. 
We say that \rocqe|B| is a branch of a tableau tree \rocqe|T| (\rocqlink{is_branch_of}) if \rocqe|B| is a path from the root of \rocqe|T| to a node with no children. 
Moreover, a formula is on a branch of \rocqe|T| (\rocqlink{is_on_branch}) if it appears in one of the labels along that branch.
We also define the following functions:
\begin{itemize}
\item \rocqe|get_context| (\rocqlink{get_context}), which accumulates the list of formulas
  along a branch,
\item {\ttfamily\small get\_all\_formulas} (\rocqlink{get_all_formulas}), which is the union of \rocqe|get_context| for all the branches of
  a tableau,
\item \rocqe|expand_tableau_branch| (\rocqlink{expand_tableau_branch}), which given a
  branch \rocqe|B|, a tableau \rocqe|T| such that \rocqe|is_branch_of B T| holds, and two
  optional list of formulas corresponding to the potential labels of the children of the
  last node of \rocqe|B|, adds the non-\rocqe|None| list of formulas as children of the
  last node of \rocqe|B|, and
\item \rocqe|replace_child| (\rocqlink{replace_child}), which replaces the last
  \rocqe|TableauTree| of a given path in a tableau tree.
\end{itemize}
A \rocqe|Tableau| (\rocqlink{Tableau}) is a pair consisting of a \rocqe|TableauTree|
together with a \rocqe|sko_record|. We define an implicit coercion from a \rocqe|Tableau| to its underlying \rocqe|TableauTree|.

\begin{figure}[t]
  \centering
  \fbox{
  \parbox{0.97\linewidth}{
    \begin{mathpar}
      \galinferrule[AlphaNegNeg]{
        \rocqm{is\_branch\_of B T} \\
        \rocqm{is\_on\_branch (Neg (Neg F)) B T}}{
        \expands{\rocqm{T}}{\rocqm{expand\_tableau\_branch (Some [F]) None B T}}
      }
      \label{rule:alpha-neg-neg}

      \galinferrule[AlphaNegOr]{
        \rocqm{is\_branch\_of B T} \\
        \rocqm{is\_on\_branch (Neg (Or F G)) B T}}{
        \expands{\rocqm{T}}{\rocqm{expand\_tableau\_branch (Some [Neg F ; Neg G]) None B T}}
      }
      \label{rule:alpha-neg-or}

      \galinferrule[BetaOr]{
        \rocqm{is\_branch\_of B T} \\
        \rocqm{is\_on\_branch (Or F G) B T}}{
        \expands{\rocqm{T}}{\rocqm{expand\_tableau\_branch (Some [F]) (Some [G]) B T}}
      }
      \label{rule:beta-or}

      \galinferrule[GammaAll]{
        \rocqm{is\_branch\_of B T} \\
        \rocqm{is\_on\_branch (All F) B T} \\
        \rocqm{x :}\ \mathcal{V}}{
        \expands{\rocqm{T}}{\rocqm{expand\_tableau\_branch (Some
            [\variableOpening{F}{0}{Free x}]) None B T}}
      }
      \label{rule:gamma-all}

      \galinferrule[DeltaNegAll]{
        \rocqm{is\_sko t (Neg (All F)) (symbols T) (fv (get\_context B T))} \\
        \rocqm{(function_symbols (get_all_formulas T)) = true} \\
        \rocqm{is\_branch\_of B T} \\
        \rocqm{is\_on\_branch (Neg (All F)) B T} \\
        \rocqm{symbs = add\_symbol (symbol t) (Neg (All F)) (symbols T)}}{
        \expands{\rocqm{T}}{%
          \rocqm{expand\_tableau\_branch (Some [\variableOpening{F}{0}{t}]) None B (T,
            symbs)}}
      }
      \label{rule:delta-neg-all}

  \end{mathpar}}}
  \caption{Expansion Rules \rocqlink{ExpansionStep}}
  \label{fig:expansion-rules}
\end{figure}

The second step of our definition consists in introducing a binary relation between tableaux
that effectively implements the tableau expansion rules---\ie all rules except the closure
rules. This relation is denoted \rocqe@T |> T'@ and described in
\cref{fig:expansion-rules}.
The most notable change w.r.t.\ the tableau rules of \cref{fig:tableau-rules} is Rule~\nameref{rule:delta-neg-all} that expects a term \rocqe|t| and uses the given
Skolemization boolean predicate to check whether \rocqe|t| is a valid Skolem function in
the context. 
In this case, the Skolem symbol of \rocqe|t| is added to the
\rocqe|sko_record| in order to be used in subsequent Skolemization steps of the
tableau. 
The other rules are implemented straightforwardly from the corresponding tableau rules. By convention, in unary rules, we extend the left child of a node.

The third step of our definition mirrors the closure rules. We say that
a tableau \rocqe|T| is \emph{closed under the substitution $\sigma$}
(\rocqlink{is_tableau_closed}) iff, for every branch \rocqe|B| of \rocqe|T|, either
\rocqe|Bot| belongs to the context of \rocqe|B|, or there are two formulas \rocqe|F| and
\rocqe|G| in the context of \rocqe|B| s.t. \rocqe|F@[sigma] = Neg G@[sigma]|.

Using these three elements, we say that the context $\Gamma$ has a tableau under the
substitution $\sigma$ (\rocqlink{hasTableau}) iff there exists a \rocqe|Sequence| (\ie{} a list of
\rocqe|Tableau|x) \rocqe|s| such that (i) the first element of \rocqe|s| is the tableau with a
single node labeled by $\Gamma$, (ii) \rocqe@s[i] |> s[i+1]@ for every \rocqe|i|, and (iii) the last element of \rocqe|s| is a closed tableau under the substitution $\sigma$.

Note that, if a sequence \rocqe|s| satisfies \rocqe@s[i] |> s[i+1]@ for every \rocqe|i|,
we call \rocqe|s| an \emph{expansion sequence}
(\rocqlink{is_expansion_sequence}). A tableau \rocqe{T} is said
  \emph{satisfiable} (\rocqlink{is_tableau_satisfiable}) if there exists a model \rocqe{M}
  such that for every free-variable environment $\mu$, there exists a branch \rocqe{B} of
  \rocqe{T} such that \rocqe@[[ M # [] # mu '|= get_context B T ]]@.

\subsection{Soundness of the Calculus}
\label{sec:soundness}

In order to use the \rocqe|hasTableau| predicate for checking proofs or in other
developments, a key property we need to show is soundness of the system:
\begin{quote}
  For every closed context $\Gamma$ and formula \rocqe|F|, if there exists $\sigma$ such that
  \rocqe|hasTableau sko (Neg F :: Gamma) sigma|, then $\validunder{\Gamma}{\rocqm{F}}$.
\end{quote}
where \rocqe|Neg F :: Gamma| denotes adding \rocqe|Neg F| to the list of formulas
\rocqe|Gamma|.

Intuitively, this follows from the fact that, for every tableau expansion rule of
\cref{fig:tableau-rules}, the tableaux \emph{before} and \emph{after} the expansion are
equisatisfiable, provided that the Skolemization is valid---\ie{} if the interpretation of $\rocqm{sko}(X_1, \ldots, X_n)$ can be assigned to an arbitrarily chosen value in the model. 
Since the last tableau of the sequence is \emph{closed}, every branch contains a contradiction and is therefore unsatisfiable.

In fact, full equisatisfiability is not required to establish soundness~\cite{MFit96}. It
suffices to show that (i) a closed tableau is unsatisfiable and (ii) the tableau expansion
rules preserve satisfiability. The first point is straightforward: the substitution
$\sigma$ can be seen as a free-variable environment under which no branch can be satisfied,
regardless of the model.

\begin{rocqlemma}[Closed Tableau is not Satisfiable][hasTableau_not_satisfiable]\label[lemma]{lem:hasTab-not-sat}
  For every model \rocqe|M|, context \rocqe|Gamma|, substitution $\sigma$ and sequence
  \rocqe|s|, if \rocqe|s| is an expansion sequence such that the first element of
  \rocqe|s| is the tableau with a single node labelled by $\Gamma$ and the last tableau of
  \rocqe|s| is closed, then no \rocqe|B| branch of the last tableau of \rocqe|s| is
  satisfied by \rocqe|M| under the substitution $\sigma$ coerced as a free-variable environment.
\end{rocqlemma}

To demonstrate the second point, we prove that the expansion rules of \cref{fig:expansion-rules} are satisfiability preserving.

\begin{rocqlemma}[Expansion Preserves Satisfiability][satisfiable_expansion_satisfiable]\label[lemma]{lem:subject-reduction}
  For every tableaux \rocqe|T| and \rocqe|T|$'$, if \rocqe|T| is satisfiable and
  \rocqe@T |> T'@, then \rocqe|T'| is satisfiable.
\end{rocqlemma}

\begin{proof}
  By case analysis over the expansion rule used. Cases \nameref{rule:alpha-neg-neg},
  \nameref{rule:alpha-neg-or}, \nameref{rule:gamma-all} are straightforward applications
  of the assumption that \rocqe|T| is satisfiable.

  \emph{Case \nameref{rule:beta-or}.} The model \rocqe|M| given by the satisfiability of \rocqe|T|
  is the one that satisfies the extended tableau.  Without loss of generality, given a free-variable environment $\mu$, we assume that the last expanded branch \rocqe|B| is the satisfiable one. Then, as \interpret{\rocqm{M}}{\cdot}{\mu}{\rocqm{Or F G}}, either
  \interpret{\rocqm{M}}{\cdot}{\mu}{\rocqm{F}}, which makes the left child of \rocqe|B|
  satisfiable, or \interpret{\rocqm{M}}{\cdot}{\mu}{\rocqm{G}}, which makes the right child
  of \rocqe|B| satisfiable.

  \setRocqFilename{Tableaux.Semantics}
  \emph{Case \nameref{rule:delta-neg-all}.} Assuming that there is a model \rocqe|M| that
  satisfies \rocqe|T|, we must construct a model that satisfies \rocqe|T| extended with
  \rocqe|Neg F{0 \to t}| on the branch \rocqe|B|. We define this model to be \rocqe|M| inside
  which we replace the interpretation function (\rocqlink{ReplacementModel}) by the one
  given by the Requirement~\ref{requirement:sko-sound} of \cref{sec:skolemization}.

  \begin{requirement*}[Requirement~\ref{requirement:sko-sound}]
    For every term \rocqe|t|, formula \rocqe|F|, function symbols \rocqe|s| and model
    \rocqe|M|, if \rocqe|t| is a valid Skolemized term, then there exists an
    interpretation function \rocqe|f : func -> list M -> M| such that:
    \begin{enumerate}[(i)]
    \item\label{item:1} for every free-variable environment $\mu$, if the function symbols
      of \rocqe|F| are a subset of \rocqe|s| and
      \rocqe@[[ M # \cdot # mu '|= Neg (All F) ]]@ then
      \rocqe@[[ ReplacementModel M f # \cdot # mu '|= Neg F{0 \to t} ]]@, and
    \item\label{item:2} for every formula \rocqe|G| and free-variable environment $\mu$, if
      the function symbols of \rocqe|G| are a subset of \rocqe|s| and if
      \rocqe@[[ M # \cdot # mu '|= G ]]@, then
      \rocqe@[[ ReplacementModel M f # \cdot # mu '|= G ]]@.
    \end{enumerate}
  \end{requirement*}

  Then, given a free-variable environment in \rocqe|ReplacementModel M f| (i.e., a
  free-variable environment in \rocqe|M|), there are two further cases. Either the
  satisfiable branch is not the one that was expanded, and by Item~(\ref{item:2}) it
  remains satisfiable in the new model; or it is the expanded branch, in which case it is
  satisfiable by the two properties of Requirement~\ref{requirement:sko-sound}.
\end{proof}

Moreover, we can show that the Requirement~\ref{requirement:sko-sound} is
fulfilled by the Skolemization strategies.

\setRocqFilename{Tableaux.Skolemization}
\begin{rocqlemma}[\dummyName][isSkolemization_OuterSkolemizationData]
  Outer Skolemization satisfies the four requirements of Skolemization (\emph{c.f.}, \cref{sec:skolemization}).
\end{rocqlemma}

\setRocqFilename{Tableaux.Semantics}
\begin{proof}
  We focus on the last requirement. First, we remark that given a free-variable
  environment $\mu$, we can write a function that returns a mere element \rocqe|c| of
  \rocqe|M| such that if \rocqe@[[ M # \cdot # mu '|= Neg (All F) ]]@, then
  \rocqe@[[ M # [c] # mu '|= Neg F ]]@ (\rocqlink{satisfy_delta}). Indeed, by using the
  full power of classical logic, we can decide whether
  \rocqe@[[ M # \cdot # mu '|= Neg (All F) ]]@ is true or false. In the former
  case, we take the element \rocqe|c| yielded by the hypothesis. In the latter
  case, we choose an arbitrary member of \rocqe|M|.

  Therefore, using the choice axiom---which is unavoidable as Skolemization is a
  function that chooses non constructively a constant value that makes the instantiated
  formula equisatisfiable w.r.t. the quantified one---we can extract this \rocqe|c| from the
  previous function. Then, given \rocqe|f : func| and \rocqe|l : list M|, the
  interpretation function acts as follows:
  \begin{itemize}
  \item if \rocqe|f| is the Skolem symbol, we update the free-variable assignment to be
    $[X_i \mapsto c_i]$, where $X_i$ is the $i$-th argument of the Skolem function and
    $c_i$ the $i$-th member of the list \rocqe|l|, and return the corresponding \rocqe|c|
    yielded by the choice function, and
  \item otherwise, return the result of the interpretation function of \rocqe|M|.
  \end{itemize}
  As the Skolem symbol is fresh, it is clear that any formula satisfied in \rocqe|M| is
  also satisfied in the new model. Moreover, \rocqe|Neg F{0 \to t}| is also satisfiable by
  commutation of variable opening and bound-variable environment and that the
  interpretation of \rocqe|t| in the new model is \rocqe|c|.
\end{proof}

\setRocqFilename{Tableaux.Skolemization}
Since the Skolem symbol is also fresh, this proof can be reused in the inner Skolemization
(\rocqlink{isSkolemization_InnerSkolemizationData}) case. The justification for pre-inner Skolemization is more involved, relying ultimately
on the fact that the Skolem symbol appears only in the formula from which it
originates. As this formula is satisfied in the model, the presence of the Skolem symbol
does not affect the satisfiability of the other formulas.

A simple corollary of \cref{lem:subject-reduction} is that an entire expansion sequence is satisfiable provided that its first tableau is satisfiable.

\setRocqFilename{Tableaux.Proofs}
\begin{rocqcorollary}[Expansion Sequence Preserves Satisfiability][satisfiable_tableau_satisfiable_expansion_sequence]\label[corollary]{cor:sequence-sat}
  For every expansion sequence \rocqe|s| and tableau \rocqe|T|, if \rocqe|s| starts by
  \rocqe|T| and \rocqe|T| is satisfiable, then \rocqe|s[i]| is satisfiable for every
  \rocqe|i|.
\end{rocqcorollary}

\begin{remark}
  Note that, using \cref{cor:sequence-sat}, it is straightforward to construct a model that
  satisfies every tableau in the sequence: it suffices to show that inverting the tableau
  rules preserves satisfiability in the same model (which is easy). For instance, this is
  the approach chosen in the literature to prove the soundness of pre-inner
  Skolemization~\cite{beckert1993delta++}.
\end{remark}

Finally, we can bring all the components together to show the soundness theorem.

\setRocqFilename{Tableaux.Proofs}

\begin{rocqtheorem}[Soundness][hasTableau_sound]
  For every substitution $\sigma$, context $\Gamma$ and formula \rocqe|F|, if $\Gamma$ and \rocqe|F| are
  closed and \rocqe|Neg F :: Gamma| has a tableau under the substitution $\sigma$, then
  $\validunder{\Gamma}{\rocqm{F}}$.
\end{rocqtheorem}

\begin{proof}
  Let \rocqe|M| be a model. Assume that \rocqe|M| satisfies \rocqe|Neg F :: Gamma| in the
  empty free-variable environment. As \rocqe|Neg F :: Gamma| is closed, \rocqe|M| actually
  satisfies \rocqe|Neg F :: Gamma| in any free-variable environment. Let \rocqe|T| be the
  last tableau of the sequence for \rocqe|Neg F :: Gamma|. By \cref{cor:sequence-sat},
  \rocqe|T| is satisfiable, \ie there exists a model \rocqe|M|$'$  s.t., for every free-variable
  environment, at least one branch of \rocqe|T| is satisfied. By \cref{lem:hasTab-not-sat}, we know that
  the substitution $\sigma$ coerced to a free-variable environment does not satisfy \rocqe|T|,
  which is a contradiction.
\end{proof}


\section{A Fully Certified Proof Checker}
\label{sec:proof-checker}

In this section, we present the different parts of the proof-checker algorithm implemented in \TableauxRocq. First, as our goal is to use this algorithm with real-world
\ac{atps}, we extend \TableauxRocq to the full \ac{fol} syntax in
\cref{sec:extended-fol}. Next, we describe the design of the proof-checking algorithm in
\cref{sec:algo-proof-check}. Finally, we prove its soundness in \cref{sec:algo-soundness}.

\subsection{Extended First-Order Logic}
\label{sec:extended-fol}

\setRocqFilename{Tableaux.ExtendedSyntax}

As first-order \ac{atps} are not restricted to the fragment of \ac{fol} used in the
previous sections, we added an extended syntax of terms (\rocqlink{ETerm}) and formulas
(\rocqlink{EForm}) to \TableauxRocq.  This extended syntax supports full first-order
logic, both in terms of syntax and semantics (\rocqlink{interpret_eterm},
\rocqlink{interpret_eform}), with string-valued variables, function symbols and predicate
symbols.

We can then easily define translation functions between \rocqe|ETerm|s and \rocqe|Term|s
(\rocqe{translate_ETerm} \rocqlink{translate_ETerm}), and between \rocqe|EForm|s and
\rocqe|Form|s (\rocqe{translate_EForm} \rocqlink{translate_EForm}), by storing bound
variables in a list \rocqe|m| and translating a standard variable to its index in
\rocqe|m| if found, or to a free variable otherwise.  With this translation suitably
defined, we obtain a correspondence between the fragment and the full semantics.

\begin{rocqlemma}[\dummyName][gen_interp_term_interp_eterm]
  For every model \rocqe|M|, extended term \rocqe|t|, list of strings \rocqe|bvs|,
  bound-variable environment \rocqe|rho| and free-variable environment \rocqe|sigma|,
  \rocqe@[[ M # rho # sigma '|= translate_ETerm bvs t]]@ is equal to
  \rocqe|interpret_eterm M (extended_env bvs rho sigma) t|.
\end{rocqlemma}

In this lemma, the \rocqe|extended_env| function maps a string \rocqe|x| to
\rocqe|rho[n]| if \rocqe|x|
has index \rocqe|n| in \rocqe|rho|, or to \rocqe|sigma(x)| if it is not found.

\begin{rocqlemma}[\dummyName][gen_translation_equivalidity]
  For every model \rocqe|M|, extended formula \rocqe|F|, list of strings \rocqe|bvs|,
  bound-variable environment \rocqe|rho| and free-variable environment \rocqe|sigma|,
  \rocqe@[[ M # rho # sigma '|= translate_EForm__aux bvs F]]@ iff
  \rocqe|interpret_eform M (extended_env bvs rho sigma) F|.
\end{rocqlemma}

Thus, the translation of any formula is valid iff the corresponding base formula is valid in the extended semantics. This allows us to show that the soundness theorem for tableaux carries over to the extended syntax.

\begin{rocqtheorem}[Soundness of the Extended Syntax][hasTableau_is_evalid]
  For every extended formula \rocqe|F|, Skolemization \rocqe|sko|, context \rocqe|Gamma|
  and substitution \rocqe|sigma|, if the translation of \rocqe|Neg F :: Gamma| is closed
  and has a tableau, then $\Gamma$ implies $\rocqm{F}$ in the extended semantics.
\end{rocqtheorem}

\subsection{Proof-Checking Algorithm}
\label{sec:algo-proof-check}

\setRocqFilename{Tableaux.Checker}

A proof-checking algorithm should be \emph{informative}, \ie it does not only check
whether a proof is valid or not, but also returns the exact error encountered on an
invalid proof.  Without this, it would be difficult for \ac{atps} developers to fix the reported
issues, which are often benign.  Hence, to enable the printing of messages, we work in the
\rocqe|Result| monad defined as
$\rocqm{Result A} \defeq \rocqm{A} \times \rocqm{list string}$, with:
\begin{itemize}
\item unit: \rocqe|ret x := (x, [])|, and
\item bind: \rocqe|r >>= f := let r' := f (fst r) in (fst r', snd r ++ snd r')|,
\end{itemize}
for which we define a notation that is similar to Haskell's \rocqe|do| block; \ie
\begin{rocq}
b <- CheckProof sko Gamma sigma T;
ret (negb b).
\end{rocq}
is syntactic sugar for
\begin{rocq}
CheckProof sko Gamma sigma T >>= fun b => ret (negb b).
\end{rocq}

In this snippet, the \rocqe|CheckProof| function (\rocqlink{CheckProof}) is the
implementation of the proof-checking algorithm defined subsequently in this section. It
takes as inputs (i) a Skolemization technique, (ii) a context \rocqe|Gamma| (here, a list
of formulas), (iii) a substitution $\sigma$, and (iv) a \rocqe|RuleTree|, which implements the
fragment set of tableaux rules (detailed in \cref{listing:extended-rules}), and returns a
\rocqe|Result bool|.

\begin{lstlisting}[caption={Rule Tree Definition \rocqlink{RuleTree}}, label={listing:extended-rules}, language=rocq]
Inductive Rule : Type :=
| AlphaNegNeg : Form -> Rule | AlphaNegOr : Form -> Rule
| BetaOr : Form -> Form -> Rule
| GammaAll : Form -> string -> Rule
| DeltaNegAll : Form -> Term -> Rule.

Inductive RuleTree : Type :=
| Leaf : Option (Form * Form) -> RuleTree
| Node : RuleTree -> Rule -> RuleTree -> RuleTree.
\end{lstlisting}

The actual algorithm (\rocqlink{CheckProof__aux}) is also parameterized by a set of
function symbols corresponding to the constant symbols of the initial context, as well as
a Skolemization record of symbols introduced by the \rocqe|DeltaNegAll| rules, kept in
the result. It is defined by induction over the \rocqe|RuleTree| argument as follows:
\begin{itemize}
\item On a \rocqe|Leaf|, there are two cases:
  \begin{itemize}
  \item If no pair of formulas is given, we search for a trivial contradiction in \rocqe|Gamma|. If found, we return \rocqe|true| with no error messages. Otherwise, we return \rocqe|false| and inform the user that \rocqe|Gamma| contains no trivial contradiction.
  \item If a pair of formulas is given, we check that both are indeed in \rocqe|Gamma| and that one of the formula is the negation of the other after applying $\sigma$. If any of these conditions fails, we return a message indicating that there are no contradictions in \rocqe|Gamma@[sigma]|.
  \end{itemize}
\item On a \rocqe|Node|, there are three cases:
  \begin{itemize}
  \item In the \rocqe|AlphaNegNeg|, \rocqe|AlphaNegOr|, and \rocqe|GammaAll| cases, we check that the given formula is in \rocqe|Gamma| and call the algorithm recursively on \rocqe|Gamma| augmented with the formulas resulting from the rule application. If any of these steps fails, we return \rocqe|false| together with the corresponding message.
  \item In the \rocqe|BetaOr| case, we first check that the target formula is in \rocqe|Gamma|. We then call the algorithm recursively on \rocqe|Gamma| augmented with the left sub-formula. If this call returns an error, we propagate it. Otherwise, we recursively call the algorithm on \rocqe|Gamma| augmented with the right sub-formula, along with the set of symbols returned by the first call, and return its result.
  \item In the \rocqe|DeltaNegAll| case, we check that the target formula is in \rocqe|Gamma| and that the given term is a valid Skolemized term w.r.t.\ the given Skolemization technique. We then recursively call the algorithm on \rocqe|Gamma| augmented with the resulting formula, adding the function symbol of the given term to the Skolemization record, and return its result.
  \end{itemize}
\end{itemize}

The \rocqe|CheckProof| function then calls \rocqe|CheckProof__aux|, with an empty
Skolemization record and the set of function symbols of the given list of formulas. Then,
it discards the record returned by \rocqe|CheckProof__aux| in the result.

\subsection{Soundness of the Algorithm}
\label{sec:algo-soundness}

Proving the algorithm sound means that, for every input proof validated by
\rocqe|CheckProof|, we must build a proof of \rocqe|hasTableau| from it.
Consequently, this proceeds in three steps: (i) show that a positive answer from a
call of \rocqe|CheckProof| implies that a sequence of tableaux can be extracted, (ii) show that
this sequence is an expansion sequence, and (iii) show that the last tableau in this sequence
is closed.

The first point can be addressed by extracting a sequence directly from a
\rocqe|RuleTree|. This is the role of the function given in
\cref{listing:ruletree-to-sequence}.

\begin{lstlisting}[caption={Translating a Rule Tree Into a Sequence of Tableaux \rocqlink{RuleTree_to_Sequence__aux}}, label={listing:ruletree-to-sequence}, language=rocq]
Fixpoint RuleTree_to_Sequence__aux (sko : Skolemization) (B : Branch) (T : Tableau) (R : RuleTree) : option Sequence := match R with
  | Leaf _ => ret [T]
  | Node R_1 r R_2 =>
      match r with
      | AlphaNegNeg F =>
         Gamma <- get_neg_neg F;
         T' <- expand_tableau_branch sko (Some (Ctx.elements Gamma)) None B T;
         s <- RuleTree_to_Sequence__aux (B ++ [Left]) T' R_1;
         ret (T :: s)
      | AlphaNegOr F =>
         Gamma <- get_neg_or F;
         T' <- expand_tableau_branch sko (Some (Ctx.elements Gamma)) None B T;
         s <- RuleTree_to_Sequence__aux (B ++ [Left]) T' R_1;
         ret (T :: s)
      | BetaOr F =>
         Gammas <- get_or F;
         T' <- expand_tableau_branch sko (Some (Ctx.elements (fst Gammas)))
                (Some (Ctx.elements (snd Gammas))) B T;
         s_1 <- RuleTree_to_Sequence__aux (B ++ [Left]) T' R_1;
         s_2 <- RuleTree_to_Sequence__aux (B ++ [Right]) (last s_1 (mkLeaf sko)) R_2;
         ret (T :: removelast s_1 ++ s_2)
      | GammaAll F x =>
         G <- get_all F;
         T' <- expand_tableau_branch sko (Some [G{0 \to Free x}]) None B T;
         s <- RuleTree_to_Sequence__aux (B ++ [Left]) T' R_1;
         ret (T :: s)
      | DeltaNegAll F t =>
         G <- get_neg_all F;
         f <- get_symbol t;
         T_0 <- expand_tableau_branch__aux (Some [G{0 \to t}]) None B T;
         let T' := {| tree := T_0; symbols := add_symbol f F (symbols T) |} in
         s <- RuleTree_to_Sequence__aux (B ++ [Left]) T' R_1;
         ret (T :: s)
      end
  end.
\end{lstlisting}

This function satisfies a few small properties. First, the extracted sequence is never empty; in fact, the first element of this sequence is always the tableau given as a parameter.

\begin{rocqlemma}[\dummyName][RuleTree_to_Sequence_hd]\label[lemma]{lem:expansion-hd}
  For every rule tree \rocqe|R|, branch \rocqe|B|, tableau \rocqe|T|, and sequence
  \rocqe|s|, if \rocqe@RuleTree_to_ Sequence__aux B T R = Some s@, then the first element of
  \rocqe|s| is \rocqe|T|.
\end{rocqlemma}

Moreover, the function only acts on the branch \rocqe|B| and its extensions; that is, given a tableau \rocqe|T|, the last element of the extracted sequence is the tableau \rocqe|T| in which the last node of the branch \rocqe|B| has been replaced by another non-empty tableau.

\begin{rocqlemma}[\dummyName][RuleTree_to_Sequence_branch]
For every rule tree \rocqe|R|, branch \rocqe|B|, tableau \rocqe|T| and sequence
\rocqe|s|, if \rocqe|B| is a branch of \rocqe|T| and
\rocqe|RuleTree_to_Sequence__aux B T R = Some s|, then there exists a non-empty tableau
tree \rocqe|T|$'$ such that
\rocqe|replace_child B T T' = Some (tree (last s (mkLeaf sko)))|.
\end{rocqlemma}

In the previous statement, \rocqe|last| is a function that returns the last element of the
sequence \rocqe|s| if it exists. Otherwise, it returns the tableau given in the second
argument, which here is a leaf---\ie an empty tableau. Both of these properties ensure that
\rocqe|RuleTree_to_Sequence__aux| returns a sequence of tableaux whenever the
\rocqe|CheckProof__aux| function returns a positive answer.

\begin{rocqlemma}[\dummyName][CheckProof_Some_RuleTree_to_Sequence_Some__aux]
  For every substitution $\sigma$, set of function symbols \rocqe|fs|, rule tree
  \rocqe|R|, branch \rocqe|B|, tableau \rocqe|T| and Skolemization records \rocqe|r| and
  \rocqe|r|$'$, if
  \rocqe@CheckProof__aux sko fs (get_context B T) sigma r R = ret {| status := true; symbs := r' |}@
  and \rocqe|B| is a branch of \rocqe|T|, then there exists a sequence of tableaux
  \rocqe|s| such that \rocqe|RuleTree_to_Sequence__aux B T R = Some s|.
\end{rocqlemma}

We can further characterize this sequence by the fact that the
Skolemization record of the last tableau of the sequence is exactly the record returned
by the proof-checking algorithm.

\begin{rocqlemma}[\dummyName][RuleTree_to_Sequence_symbols]
For every substitution $\sigma$, set of function symbols \rocqe|fs|, rule tree \rocqe|R|, branch \rocqe|B|,
  tableau \rocqe|T| and Skolemization record \rocqe|r|,
  if \rocqe|B| is a branch of \rocqe|T|,
   \rocqe@CheckProof__aux sko fs (get_context B T) sigma (symbols T) R = ret {| status := true; symbs := r |}@,
  and 
  \rocqe@RuleTree_to_Sequence__aux B T R = Some s@, then \rocqe|r = symbols (last s (mkLeaf sko))|.
\end{rocqlemma}

Recall that the set of function symbols provided to \rocqe|CheckProof__aux| is the constant set of function symbols from the initial context. We use this set together with the Skolemization record of the given tableau to check whether a term is a valid Skolemization. While this avoids having to keep track of all the function symbols occurring in the tableau, it prevents us from directly satisfying Rule~\nameref{rule:delta-neg-all}, which requires the function symbols of all formulas in the tableau.
However, intuitively, no expansion step adds a function symbol to the tableau, except
Rule~\nameref{rule:delta-neg-all}, which records the symbol in the Skolemization
record. Consequently, the set of function symbols of the whole tableau can be
reconstructed from the initial set of function symbols together with the generated Skolem
symbols.  We say that a tableau \rocqe|T| \emph{preserves the function symbols} of
\rocqe|s|
(\setRocqFilename{Tableaux.Proofs}\rocqlink{preserves_function_symbols}\setRocqFilename{Tableaux.Checker})
if the function symbols of the tableau \rocqe|T| are included in the union of \rocqe|s|
and the Skolemization record of \rocqe|T|. Note that we cannot obtain a stronger property,
since there is no guarantee that a newly generated Skolem symbol appears in a formula (\eg
when the quantified variable does not occur in the subformula).

Assuming the preservation of function symbols allows us to show that if the proof-checking
algorithm returns a positive answer and \rocqe|s| is the extracted sequence, then
\rocqe@s[0] |> s[1]@.

\begin{rocqlemma}[\dummyName][RuleTree_to_Sequence_snd_expansion]\label[lemma]{lem:snd-expansion}
  For every substitution $\sigma$, set of function symbols \rocqe|fs|, rule tree \rocqe|R|,
  branch \rocqe|B|, tableaux \rocqe|T|, \rocqe|T|$'$, Skolemization record \rocqe|r|, and
  sequence \rocqe|s|, if \rocqe|B| is a branch of \rocqe|T|, the function symbols
  \rocqe|fs| are preserved by \rocqe|T|,
  \rocqe@CheckProof__aux sko fs (get_context B T) sigma (symbols T) R = ret {| status := true; symbs := r |}@
  and \rocqe|RuleTree_to_Sequence__aux B T R = Some s|, such that \rocqe|s[1] = T|$'$, then
  \rocqe@T |> T@'.
\end{rocqlemma}
 
By induction, we recover that under the same assumptions, \rocqe|s| is an expansion
sequence.

\begin{rocqcorollary}[Proof Checker Produces an Expansion Sequence][CheckProof_Some_RuleTree_to_Sequence_is_expansion_sequence]
    For every substitution $\sigma$, rule tree \rocqe|R|, branch \rocqe|B|,
  tableaux \rocqe|T|, \rocqe|T|$'$, Skolemization record \rocqe|r| and sequence \rocqe|s|,
  if \rocqe|B| is a branch of \rocqe|T|, the function symbols
  \rocqe|fs| are preserved by \rocqe|T|,
   \rocqe@CheckProof__aux sko (get_context B T) sigma (symbols T) R = ret {| status := true; symbs := r |}@
   and \rocqe|RuleTree_to_Sequence__aux B T R = Some s|, then \rocqe|s| is an expansion
   sequence.
\end{rocqcorollary}

\begin{proof}
  By induction on the length of the sequence. As it cannot be empty by
  \cref{lem:expansion-hd}, we only need to show the property for the inductive
  case. Because of \cref{lem:snd-expansion}, all of the cases except the one for the Rule
  \nameref{rule:beta-or} are straightforward. For the branching rule, we recover two
  sequences \rocqe|s_1| and \rocqe|s_2| by induction hypothesis, the first one starting
  from \rocqe|T| and the second one starting from the last tableau of \rocqe|s_1|. Recall
  that the sequence yielded here is \rocqe|T :: removelast s_1 ++ s_2|. To show that this
  is an expansion sequence, consider an index $i \in \mathbb{N}$:
  \begin{itemize}
  \item If $i < \#|\rocqm{s}_1| - 1$, then if $i = 0$, we directly conclude using
    \cref{lem:snd-expansion}. Otherwise, we use the fact that \rocqe|s_1| is an expansion
    sequence.
  \item If $i = \#|\rocqm{s}_1| - 1$, then we need to show that the last tableau of
    \rocqe|removelast s_1| reduces to the first element of \rocqe|s_2|. By
    \cref{lem:expansion-hd}, the first element of \rocqe|s_2| is the last tableau of
    \rocqe|s_1|, and we conclude this case using the fact that \rocqe|s_1| is an expansion
    sequence.
  \item Otherwise, we need to show that there is a reduction between two elements of
    \rocqe|s_2|, which is directly given by the induction hypothesis.
  \end{itemize}
\end{proof}

Combining the previous statements is enough to show that if the proof-checking algorithm
returns a positive answer, then we can extract an expansion sequence from the rule
tree. We can finally show that the last element of this sequence is closed.

\begin{rocqlemma}[Proof Checker Validates only Closed Tableaux][CheckProof_Some_RuleTree_to_Sequence_closed]\label[lemma]{lem:seq-closed}
  For every substitution $\sigma$, rule tree \rocqe|R|, branches \rocqe|B| and \rocqe|B|$'$,
  tableau \rocqe|T|, Skolemization record \rocqe|r|, and sequence \rocqe|s|,
  if \rocqe|B| is a branch of \rocqe|T|,
   \rocqe@CheckProof__aux sko (get_context B T) sigma (symbols T) R = ret {| status := true; symbs := r |}@
   , \rocqe|B ++ B|$'$ is a branch of the last tableau
  of \rocqe|s| and \rocqe|RuleTree_to_Sequence__aux B T R = Some s|, then the branch
  \rocqe|B ++ B|$'$ is closed in the last tableau of \rocqe|s|.
\end{rocqlemma}

Since the algorithm is initiated on the empty branch, \ie an empty list of branching steps, every branch in the final tableau of the sequence is an extension of this initial branch. Consequently, if the sequence begins with a tableau consisting of a single node labeled by $\Gamma$, we can establish the closure of the final tableau by using the previous lemma. Furthermore, the algorithm is invoked on a tableau whose only node is labeled by $\Gamma$ and whose set of symbols is empty, ensuring that it preserves function symbols. From these two facts, we deduce the soundness theorem.

\begin{rocqtheorem}[Soundness of the Proof Checker][CheckProof_sound]
  For every $\Gamma$ list of formulas, $\sigma$ substitution and \rocqe|R| rule tree, if
  \rocqe|CheckProof sko Gamma sigma R = ret true| then
  \rocqe|hasTableau sko Gamma sigma|.
\end{rocqtheorem}

This allows us devise a \rocqe|tableaux| tactic that, given a list
of formulas $\Gamma$ together with a rule tree $R$ and a substitution $\sigma$, decide wether $(R, \sigma)$ is a valid tableau if the root has context $\Gamma$. It suffices to call \rocqe|CheckProof_sound| and to let \Rocq
compute the result of \rocqe|CheckProof sko Gamma sigma R|, effectively providing a fully
certified tableau proof checker.

\section{Experiments}
\label{sec:comparison-benchs}
This section presents an experimental evaluation of \TableauxRocq. We compare the proof-checking time of \TableauxRocq's certificates against that of the \Rocq terms produced by \goeland, on both a set of problems from the \tptp  library~\cite{Sut24} and on a family of problems specifically designed to stress
Skolemization.

Experiments were performed with the \goeland{} theorem prover on the Grid'5000
platform~\cite{grid5000}, using an Intel Xeon Gold 5220 CPU (Cascade Lake-SP, x86\_64,
2.20\,GHz, 18 cores, 96\,GiB RAM). \shepherd{The modified version of \goeland, the command lines
used for all experiments, as well as the complete benchmark set are provided in the
related Zenodo archive~\cite{zenodo_sm}.}

\para{Deskolemization}
\goeland{} is a proof-producing ATP able to export proofs in \Rocq (referred to as \Rocq terms),
\sctptp~\cite{guilloud2025interoperability}, \lisa~\cite{guilloud2023lisa}, and \lp~\cite{lambdapi}. In most of these target formats, proof certificates are obtained via an intermediate deskolemization phase~\cite{LPAR2024:Generic_Deskolemization_Strategy,hermant2013syntactic,bibel82atp}.
This transformation turns a \goeland{} proof into a \textsf{GS3}~\cite{troelstra2000basic} proof,
a sequent-based calculus closely related to tableaux but restricted to
outer Skolemization.
As a result, proofs produced using non-outer Skolemization (\eg{} \cite{hahnle1994delta+,beckert1993delta++,baaz1995delta*,cantone98delta**,giese1999deltaepsilon}) must be transformed,
introducing additional computational cost and increasing proof size.

\TableauxRocq avoids this limitation entirely: since it natively supports arbitrary
Skolemization strategies and substitutions, the translation directly follows the structure
of the proof without any prior transformation. In our experiments, we exploit this by
comparing the two output pipelines on proofs using the same Skolemization method:
\goeland{} \Rocq terms (after deskolemization in the inner case) versus its \TableauxRocq
output.

\para{On the \tptp Library} \shepherd{We used the \tptp library v9.2.1, from which we extracted first-order
(FOF) theorems without equality, resulting in $797$ problems.  
All
selected problems are first run with \goeland in outer and inner
Skolemization, with a time limit of $300$ seconds per problem. Of these, $172$ (resp.\ $177$) 
problems were successfully proved in outer (resp.\ inner) Skolemization and retained for the remainder of the experiment. For 
each retained problem, \goeland{} is run again to generate, in each Skolemization mode: 
(i) \Rocq terms---via deskolemization for inner Skolemization, directly for outer---and 
(ii) \TableauxRocq output. For each problem, we measure the \Rocq type-checking time, 
plus the deskolemization time where applicable.
}

\shepherd{
On this benchmark set, using inner Skolemization, the median checking time of 
\Rocq terms is $0.15$s, while the median checking time of the \de in 
\TableauxRocq is $0.45$s. On these problems, \Rocq is typically faster, as its 
kernel checks proofs by directly executing \textsf{OCaml} code, whereas 
\TableauxRocq proof-checking algorithm runs through \Rocq's \rocqe|vm_compute| 
tactic. In addition, this benchmark makes very little use 
of inner Skolemization: $171$ out of $177$ problems have identical branch counts 
in both proof formats, leaving no structural advantage. As expected, \TableauxRocq incurs a 
moderate overhead, but both checking times remain practical.
}

\shepherd{The picture changes for the \texttt{LCL686}\footnote{\href{https://tptp.org/cgi-bin/SeeTPTP?Category=Problems&Domain=LCL&File=LCL686+1.001.p}{LC686+1.001.p}, \href{https://tptp.org/cgi-bin/SeeTPTP?Category=Problems&Domain=LCL&File=LCL686+1.005.p}{LC686+1.005.p}, \href{https://tptp.org/cgi-bin/SeeTPTP?Category=Problems&Domain=LCL&File=LCL686+1.010.p}{LC686+1.010.p}, \href{https://tptp.org/cgi-bin/SeeTPTP?Category=Problems&Domain=LCL&File=LCL686+1.015.p}{LC686+1.015.p} and \href{https://tptp.org/cgi-bin/SeeTPTP?Category=Problems&Domain=LCL&File=LCL686+1.020.p}{LC686+1.020.p}.} problem family, which makes 
significant use of Skolemization. There, \TableauxRocq becomes substantially 
faster as problem size grows: at size $5$, \Rocq requires $6.2$s (+ $0.4$s 
deskolemization) against $3.2$s for \TableauxRocq; at size $15$, the gap widens 
to $103.6$s (+ $9.2$s) against $28.5$s. This confirms that \TableauxRocq's 
advantage over the \Rocq output is most pronounced precisely when Skolemization 
plays a significant role in the proof.}

\shepherd{Using outer Skolemization, \Rocq terms require no deskolemization, making this 
a direct comparison of the two execution models. On this benchmark set, the median 
checking time of \Rocq terms is $0.14$s, while the median checking time of the 
\de in \TableauxRocq is $0.42$s. \Rocq is faster on $169$ out of 
$172$ problems, consistently by a factor of roughly $3$, illustrating the 
systematic overhead of \rocqe|vm_compute| over \Rocq's \textsf{OCaml} 
kernel.}

\shepherd{Interestingly, the \texttt{LCL686} family again sees \TableauxRocq become faster 
at larger sizes ($1.2\times$ at size $5$, $1.5\times$ at sizes $10$ and $15$), 
even without any deskolemization overhead. This shows that the gain of \TableauxRocq 
on complex proofs is not solely due to deskolemization: as proof 
complexity grows, \Rocq per-branch type checking cost increases, 
whereas \TableauxRocq's structural checker scales more efficiently. Comparing with 
the inner Skolemization results, the gap at size $15$ is $101.8$s vs $67.1$s in outer Skolemization, 
against $112.8$s (including deskolemization) vs $28.5$s in inner mode. This shows 
that deskolemization accounts for roughly half of \TableauxRocq advantage in 
the inner case, with the other half coming from simpler proof term structure.
}



\begin{figure}[t]
\centering
\begin{tikzpicture}
\begin{axis}[
    ymode=log,
    width=\linewidth-5pt,
    height=6cm,
    xlabel={Problem size $n$},
    ylabel={Proof checking time (s)},
    xmin=1, xmax=25,
    xtick={1, 5,10,15,20,25},
    ymajorgrids=true,
    legend style={at={(0.55,0.95)},anchor=north west}
]

\addplot[
    solid,
    mark=o,
    mark options={solid},
    blue
]
coordinates {
    (1,0.13) 
    (2,0.15) 
    (3,0.22) 
    (4,0.44) 
    (5,2.03)
    (6,15.06)
    (7,133.12)
};
\addlegendentry{\Rocq{}}

\addplot[
    dashed,
    mark=triangle*,
    mark options={solid},
    green
]
coordinates {
    (1,0.13) 
    (2,0.15) 
    (3,0.22) 
    (4,0.48) 
    (5,2.63)
    (6,20.26)
    (7,178.22)
};
\addlegendentry{\Rocq{} + deskolemization}

\addplot[
    dotted,
    mark=square*,
    mark options={solid},
    red
]
coordinates {
    (1,0.40) 
    (2,0.40) 
    (3,0.43) 
    (4,0.48) 
    (5,0.49)
    (6,0.55)
    (7,0.55)
    (8,0.57)
    (9,0.57)
    (10,0.63)
    (11,0.74)
    (12,0.78)
    (13,0.94)
    (14,0.96)
    (15,1.12)
    (16,1.25)
    (17,1.37)
    (18,1.38)
    (19,1.65)
    (20,1.65)
    (21,1.89)
    (22,2.05)
    (23,2.16)
    (24,2.31)
    (25,2.51)
};
\addlegendentry{\TableauxRocq{}}


\end{axis}
\end{tikzpicture}
\caption{Proof Checking Time for the Skolemization Scaling Benchmarks}
\label{fig:scaling}
\end{figure}
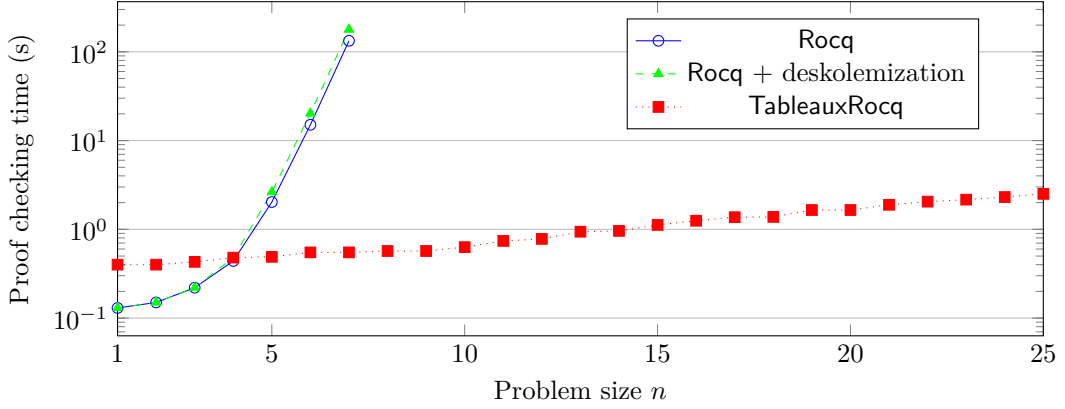

\para{On the Skolemized Problems} We also perform scaling experiments on a family of
problems with the following structure:
\begin{equation*}
  \Phi_n \defeq \forall X_1, ..., X_n \, , \, \bigvee_{i = 1}^{n} (\neg P_i(X_i) \land \exists Y_i, \,  P_i(Y_i))
\end{equation*}
These problems are specifically designed to stress Skolemization. In particular, they
typically require extensive reintroduction of Skolem terms in the outer Skolemization
mode---and thus in the deskolemization process---whereas the inner Skolemization mode leads to more straightforward proofs. 

\shepherd{
We generate instances for \( 1 \leq n \leq 25 \) and measure 
the proof-checking time for both translations, as well as the deskolemization time 
for the \Rocq output, with a time limit of 300 seconds per run. Results are shown 
in \cref{fig:scaling} in logarithmic scale. While inner Skolemization successfully 
produced proofs for all instances up to \( n = 25 \), the deskolemization algorithm 
could only handle up to \( n = 7 \), as the exponential blowup in proof size caused it 
to exceed the time limit. Indeed, the deskolemization time grows with proof size, 
remaining reasonable up to $n = 5$ ($0.6$s), but rising to $5.2$s at $n = 6$ and 
exceeding $45$s at $n = 7$, severely impacting proof generation.}

\shepherd{
Moreover, as we can see in \cref{fig:scaling}, the
checking time of the (deskolemized) \Rocq terms starts to exhibit
exponential growth, while the \TableauxRocq output displays a linear increase.
For small instances ($n \leq 3$), \Rocq terms are faster to type-check; 
the two approaches become comparable at $n = 4$ ($0.44$s + $0.04$s deskolemization 
for \Rocq terms, $0.49$s for \TableauxRocq). From $n = 5$ onward, \TableauxRocq 
becomes significantly faster: at $n = 7$, \Rocq terms require $133.1$s 
(+ $45.1$s deskolemization) to type-check, against only $0.55$s for \TableauxRocq. 
The curve further shows that \TableauxRocq scales well beyond that point, requiring 
only $2.78$s at $n = 25$. Note that our experiments were limited by \goeland's 
proof generation time, not by \TableauxRocq's proof-checking time.
}

\para{Overall Observations} It is important to note that the two checking pipelines rely
on different execution models, which introduces a systematic performance
penalty for \TableauxRocq{}, regardless of proof structure. Despite this,
\TableauxRocq{} achieves comparable checking performance on the \tptp benchmarks,
and significantly better scalability on Skolemization-based problems. Beyond performance
considerations, \TableauxRocq{} is suitable for prover integration. The tableau-based
format allows proofs to be naturally expressed in a form that closely matches the
structure of tableaux proofs, making it easier for \ac{atps} to output such
certificates. Moreover, provers using optimized Skolemization strategies do not need to
implement a dedicated deskolemization
phase: the benefits of optimized Skolemization during proof search outweigh the (slight)
additional checking cost introduced by the \TableauxRocq{} translation.

As a side note, while running experiments, some of the proofs generated by \goeland in the 
\TableauxRocq format were rejected by the checker. Upon investigation, these rejections 
exhibited actual bugs in the \goeland proof output, illustrating how \TableauxRocq's 
certified checker can serve as a debugging tool for \ac{atps}.


\section{Related Work}
\label{sec:related-work}

As far as the authors know, free-variable tableaux systems have not been formalized
before. In the literature, efforts have mostly focused on sequent-style calculi, which are
useful to study proof-theoretic properties, or on resolution-based systems.

\para{Formalization of Tableaux Calculi} Formal study of tableaux calculi are quite
rare. In first-order logic, From, Schlichtkrull and
Villadsen~\cite{DBLP:journals/logcom/FromSV23} relie on the ground version of the calculus
to prove properties of a sequent-style system. In a more general setting, Blanchette,
Popescu and Traytel~\cite{DBLP:journals/afp/Blanchette0T14} propose an abstract proof of
completeness in \isabelle, which can be instantiated in many proof systems for first-order
logic, in particular with ground tableaux. Tableaux have also been studied in a hybrid
logic by From~\cite{DBLP:conf/types/From20}, but still without free variables.

\para{Other Proof Systems} The \Rocq{} library of undecidability
proofs~\cite{forster2020coq} includes a natural deduction system---proved sound---where it is
shown that one cannot decide the validity of a formula. Clausified systems have also been
formalized, mostly in \isabelle. For instance, the resolution procedure has been formally
proved sound and complete by Schlichtkrull~\cite{DBLP:journals/jar/Schlichtkrull18}. In
the same vein, Waldmann et al.~\cite{DBLP:journals/jar/WaldmannTRB22} gave an abstract
specification of \ac{atps} using optimized saturation calculi and prove their
completeness. Building on that, Bergeron, Krasnopol and
Tourret~\cite{bergeron_et_al:LIPIcs.ITP.2025.22} formalized clause splitting, a more
advanced technique for saturation-based theorem provers. In another register,
Schlichtkrull, Blanchette and Traytel~\cite{schlichtkrull2019verified} developed a
resolution-based fully certified ATP\@, while Fleury, Blanchette and
Lammich~\cite{fleury2018verified} provided a verified SAT solver in Imperative HOL\@.

\section{Conclusion and Future Work}
\label{sec:conclusion}

We have introduced the foundations of the \TableauxRocq library, a \de of
free-variable tableaux in \Rocq that also serves as a certified proof checker for
tableaux \ac{atps}. \TableauxRocq proves that the tableau method is sound for first-order
logic, and allows for a seamless proof output from \ac{atps}. However,
\TableauxRocq is currently restricted to first-order logic without equality, leaving many avenues for future work.

A first axis is to improve the supported logic. In particular, we aim to add equational
reasoning, as most \tptp problems involve equality and are therefore not currently
certifiable by \TableauxRocq. This would enable broader experimental evaluation of
\TableauxRocq's proof-checking performance. We also plan to integrate additional
Skolemization strategies beyond inner and outer Skolemization, and to study whether a
modular framework \emph{à la} Cantone and Nicolosi-Asmundo~\cite{cantone2007sound} can be incorporated in order to ease the development of new Skolemization methods. More generally, as noted in the introduction, many tableaux provers are not based on first-order logic. Consequently, we want to extend \TableauxRocq to support multiple logics, possibly using a \emph{Coq à la Carte}~\cite{DBLP:conf/cpp/0002S20} approach to enable sharing different parts of the proofs in multiple logics.
Finally, we aim to show completeness of the system.


A second direction concerns the proof checker itself. One of our next goals is to setup an
automated extraction of \TableauxRocq's proof checker that parses a \tptp-style proof,
namely \sctptp~\cite{guilloud2025interoperability}. This serves two purposes. First, it
would make it possible to provide a standalone executable containing the fully certified
proof checker. Second, we would not have to rely on \Rocq's \rocqe|vm_compute| tactic,
enabling more accurate measurements of \TableauxRocq's proof-checking performance and
a fair(er) comparison w.r.t. the outputs directly targeting \Rocq
  terms. Ultimately, this could support the development of a
hammer-like~\cite{DBLP:journals/jar/CzajkaK18,DBLP:conf/cade/BohmeN10} tactic
\rocqe|tableaux| in \Rocq, enabling automated calls to external tableau provers together
with certified proof reconstruction.

\section*{Declaration of AI Use}

We declare that no GenAI (LLMs) was used in the writing of this paper, in the
development of the customized \goeland version and in the development of the \TableauxRocq
library.


\bibliography{biblio}

\end{document}